\documentclass[fleqn,twoside]{article}
\usepackage[headings]{espcrc2}
\usepackage{graphicx}
\usepackage{bm}

\readRCS
$Id: espcrc2.tex,v 1.2 2004/02/24 11:22:11 spepping Exp $
\usepackage{graphicx}
\usepackage[figuresright]{rotating}

\newcommand{\AmS}{{\protect\the\textfont2
  A\kern-.1667em\lower.5ex\hbox{M}\kern-.125emS}}
\newcommand{\order}[1]{{\mathcal O}\left(#1\right)}

\title{\vspace{-10mm}
  {\small $\phantom{.}$ \hfill ALBERTA-THY-09-10\\[-2mm]
  $\phantom{.}$ \hfill CERN-PH-2010-155}\\
  Magnetic moment of a bound electron}

\author{Andrzej Czarnecki\address[UofA]{Department of Physics, 
University of Alberta, Edmonton, Alberta, Canada T6G 2G7
}\address[CRN]{CERN, Theory Division, CH-1211 Geneva 23, Switzerland},
Matthew Dowling\addressmark[UofA],
        Jorge Mond\'ejar\addressmark[UofA], and
        Jan H.~Piclum\addressmark[UofA]}
       
\runtitle{Magnetic moment of a bound electron}
\runauthor{A. Czarnecki}

\begin{document}

\begin{abstract}
  Theoretical predictions underlying determinations of the fine
  structure constant $\alpha$ and the electron-to-proton mass ratio
  $m_e/m_p$ are reviewed, with the emphasis on the bound electron
  magnetic anomaly $g-2$.  The theory of the
  interaction of hydrogen-like ions with a magnetic field is discussed.  The status of
  efforts aimed at the determination of $\order{\alpha(Z\alpha)^5}$
  and $\order{\alpha^2(Z\alpha)^5}$ corrections to the $g$ factor is
  presented.  The reevaluation of analogous corrections to the Lamb
  shift and the hyperfine splitting is summarized.  \vspace{1pc}
\end{abstract}

\maketitle

\section{Introduction}
The world of our everyday experience is largely controlled by two
dimensionless fundamental quantities: the fine structure constant 
$\alpha \simeq 1/137$ and the electron-to-proton mass ratio $m_e/m_p
\simeq 1/1836$.  The former describes the strength of the
electromagnetic interaction.  The two together relate the energy of
atomic excitations to the bulk of the atomic mass.  

Among the myriads of phenomena governed by these two constants, only a
handful are suitable for their precise determination.  Two
characteristics are required: they have to be amenable to measurements
and they must be theoretically very well understood.  Most of such
systems consist of only one, two, or three bodies.  

In this talk we focus on the theoretical aspects of 
hydrogen-like ions.  Measurements of their
interaction with the magnetic field are a crucial ingredient in the
best determination of $m_e/m_p$, and indirectly help to determine
$\alpha$ with the second-best accuracy.  

We first briefly review the present knowledge of both constants, to
illustrate the level of accuracy already achieved.  A system of
ultimate simplicity is a single electron.  It is studied with the
Penning trap where it forms what is sometimes called a geonium atom
\cite{Brown:1985rh}.  The spectrum of lowest lying states is sensitive
to the electron anomalous magnetic moment, $g-2$, and thus to $\alpha$
through the theoretical expression
  \begin{eqnarray}
    \label{eq:g-2}
a 
\equiv {g-2\over 2} 
= 1 +\sum_{i=1}^\infty C_i \left({\alpha\over \pi}\right)^i
+\Delta a,
  \end{eqnarray}
where $C_i$ are the coefficients determined from an $i$-loop
correction to the interaction of an electron with a weak homogeneous
magnetic field, calculated in quantum
electrodynamics 
(QED); and
$\Delta a$ are small corrections due to weak interactions and hadronic
effects. The QED effects are now being studied at five-loops
\cite{kinoshitaLeptMomBook,laportaLeptMomBook}. Indeed, the missing
information about $C_5$ is the leading error in the present best value
of $\alpha$ \cite{Hanneke:2008tm},
\begin{equation}
\alpha^{-1}(g-2) = 137.035\,999\,084(39)_{\mathrm{th}}(33)_{\mathrm{exp}}.
\end{equation}
The relative uncertainty of this evaluation is 0.37 part per billion (ppb).

The Penning trap can also be used to measure $m_e/m_p$ by comparing
the cyclotron frequency $\omega_c=eB/m$ of an electron with that of a
proton.  The best  value obtained in this way is accurate to 2.2 ppb 
\cite{vanDyck},
\begin{equation}
{m_p \over m_e}(\mathrm{free }\,\,e^-) = 1836.152\,666\,5(40).
\label{eq:vanDyck}
\end{equation}
The systematic limitation of this determination is the relativistic
shift of the electron mass due to thermal excitations of the
magnetron motion of the electron.  For this reason, studies of
electrons bound in a heavier ion, that will be discussed below, are
superior: the thermal velocity of such a heavy system is eliminated as
the source of relevant errors. 

The second-best method for finding $\alpha$ is based on the excellent
knowledge (to within 0.007 ppb) of the Rydberg constant  \cite{Mohr:2008zz},
\begin{equation}
R_\infty \equiv 
{\alpha^2 m_e c\over 2h} = 10\, 973\, 731.568\, 527(73)/\mathrm{m},
\end{equation}
where $h$ denotes the Planck constant and $c$ the velocity of light.
The latter being an exact number, $\alpha$ can be determined if the
ratio of the electron mass to the Planck constant is measured.  Such a
direct measurement is difficult, but it is possible to measure the
ratio of an atom mass $m_\mathrm{atom}$ to $h$ for atoms such as
cesium and rubidium \cite{Cadoret:2008st,Wicht2002}.  Then, if $m_e/m_p$ and
$m_p/m_{\mathrm{atom}}$ are known, one can determine $\alpha$.
(This underscores again the importance of $m_e/m_p$.)  The best value
obtained in this way, with rubidium atoms, is
\begin{equation}
\alpha^{-1}(\mathrm{Rb}) = 137.035\,999\,450(620),
\end{equation}
a 4.6 ppb determination.  Efforts to reduce this error to about 1 ppb
are being undertaken \cite{Guellati2010}.

A macroscopic effect that  competes with the atomic scale
phenomena in the possible accuracy of determining $\alpha$ is the
quantum Hall effect \cite{jeffery97}, giving the value
\begin{equation}
\alpha^{-1}(\mathrm{QHE}) = 137.036\,003\,700(3300),
\end{equation}
whose uncertainty is 23 ppb.

The last method for $\alpha$ that we want to mention here is based on
the comparison of the measurement \cite{PhysRevA.79.060503} of the
fine structure of helium with QED predictions
\cite{PhysRevLett.104.070403}, having at present a 31 ppb uncertainty,
\begin{equation}
\alpha^{-1}(\mathrm{He}) =
137.036\,001\,100(3900)_{\mathrm{th}}(1600)_{\mathrm{exp}}. 
\end{equation}
%Like in the case of $g-2$, the error is dominated by the theory.  

In addition to the two methods of determining the electron-to-proton
mass ratio that were already mentioned, there are also ongoing efforts
employing simple molecules.  Rotational and vibrational excitations
have energies proportional to $m_e/M$ and $\sqrt{m_e/M}$, respectively.
Here $M$ denotes the mass of the nuclei.  Thus, measurements of
rovibrational spectra of systems such as ions $H_2^+$ and $HD^+$
directly access $m_e/m_p$ \cite{PhysRevLett.36.1488}. This approach
depends of course on the accurate theoretical description of the ion. 
The current status of experimental and theoretical studies is
described in 
\cite{PhysRevA.79.012501,schneiderl2010}.

\section{\boldmath Bound electron $g$ factor}
\subsection{Binding effects in \boldmath $g-2$}
We now focus on one system: an electron bound in a hydrogen-like ion.
We discuss its theoretical description, explain how the ion is used to
determine the electron-to-proton mass ratio, and discuss prospects for
improvements of the theoretical predictions. 
\begin{figure}[ht]
  \includegraphics[width=\columnwidth]{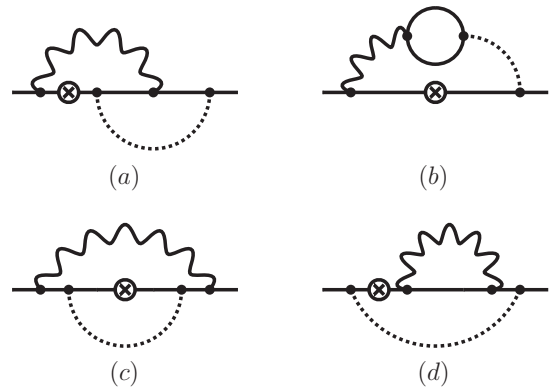}
  \caption{\label{fig::gbound} Sample diagrams for the $g$ factor of a
    bound electron at order $\alpha (Z\alpha)^5$. Solid and wavy lines
    denote electrons and photons. The dotted lines denote the
    interaction with the nucleus which is described in
    Sec.~\ref{sec::method}. The interaction vertex of the external
    photon is denoted with the symbol $\otimes$.}
\end{figure}

%\section{Lowest order}

Consider an ion consisting of a nucleus with zero spin and a single
electron. We will be interested here in the ground state of this ion.
Since the nucleus has no magnetic moment, the magnetic interaction
of this system is very similar to that of a free electron, modulo
small binding effects. We define the bound-electron gyromagnetic ratio
$g$ by the correction to the energy of the ion, linear in the magnetic
field $\bm{B}$. In the ground state, the whole magnetic moment
$\bm{\mu}$ of the ion is due to the electron's spin $\bm{s}$, so
the energy shift is\begin{equation}
\delta E=-\left\langle \bm{\mu\cdot\bm{B}}\right\rangle 
\equiv g\frac{e}{2m_e}\left\langle \bm{s}\cdot\bm{\bm{B}}\right\rangle \,,
\label{eq:gDef}\end{equation}
where $-e$ is the electron's charge. The
expectation value is to be calculated with the wave function of the
ion. 

In the lowest order, neglecting all effects except the Coulomb field,
$g$ was calculated \cite{Breit28} already in 1928, in one of the
first applications of the Dirac equation \cite{Dirac:1928hu}. To
illustrate the influence of binding effects, we
reproduce here that result, treating the interaction with the external
magnetic field as a perturbation. A uniform field is described by
the vector potential $\bm{A}=\frac{1}{2}\bm{\bm{B}}\times\bm{r}$,
and the correction to the energy is\begin{equation}
\delta E\left(\mathrm{Coulomb}\right)
=e\int d^{3}r\bar{\psi}\bm{A}\cdot\bm{\gamma}\psi\label{eq:Agamma}\,,\end{equation}
where $\psi$ is the solution of the Dirac equation with the Coulomb
potential $V\left(r\right)=-\frac{Z\alpha}{r}$. Here $Z$ is the
number of protons in the ion's nucleus and $\bm{r}$ describes the position
of the electron relative to the nucleus (we assume the nucleus to
be very heavy and motionless).
%, and $\alpha\simeq1/137.036$ is the
%fine structure constant. 
The ground state wave function can be written
in the form \cite{Schwinger:II}
%\cite{:II}
%
\begin{eqnarray}
\psi\left(\bm{r}\right) & = & f\left(r\right)
\left(1+i\gamma\gamma^{5}\bm{\Sigma}\cdot\hat{r}\right)v\,,
\label{eq:DiracWF}\\
f\left(r\right) & \equiv & Nr^{-Z\alpha\gamma}\exp\left(-Z\alpha
  m_er\right)\,,
\nonumber \\
\gamma & \equiv & \frac{1-\sqrt{1-\left(Z\alpha\right)^{2}}}
{Z\alpha}\simeq\frac{Z\alpha}{2}\,,
\nonumber \end{eqnarray}
where $v$ is a spinor such that $v^{\dagger}v=1$ and $N$ is a normalization
constant such that
\begin{equation}
\left(1+\gamma^{2}\right)\int d^{3}rf^{2}=1.
\label{eq:normalization}\end{equation}
$N$ approximately equals the value of the non-relativistic Coulomb
wave function at the origin, $N^{2}\simeq\frac{\left(Z\alpha m_e\right)^{3}}{\pi}$.
In the standard (Dirac) representation, $\bm{\Sigma}=\left(\begin{array}{cc}
\bm{\sigma} & 0\\
0 & \bm{\sigma}\end{array}\right)$ and $\gamma^{5}=\left(\begin{array}{cc}
0 & 1\\
1 & 0\end{array}\right)$. With this wave function, we find
\begin{equation}
\delta
E=\frac{8\pi}{3}e\gamma\left(v^{*}\bm{\bm{B}}\cdot\bm{\sigma}v\right)
\int_{0}^{\infty}drr^{3}f^{2}\left(r\right).
\label{eq:deltaEintegral}\end{equation}
Radial integrals of this type are
\begin{eqnarray}
\lefteqn{\int_{0}^{\infty}drr^{n}f^{2}\left(r\right)}&& \nonumber\\
&=&\frac{\left(2Z\alpha
    m_e\right)^{2-n}\Gamma\left(n+1-2Z\alpha\gamma\right)}
{4\pi\left(1+\gamma^{2}\right)\Gamma\left(3-2Z\alpha\gamma\right)}\,,
\label{eq:radialInt}
\end{eqnarray}
leading to 
\begin{equation}
g\left(\mathrm{Coulomb}\right)=
\frac{2}{3}\left(1+2\sqrt{1-\left(Z\alpha\right)^{2}}\right),
\label{eq:gCoulomb}\end{equation}
the result first obtained by Breit \cite{Breit28}.

\subsection{Mixed binding-selfinteraction effects}
In view of recent precise measurements with hydrogen-like ions, a
variety of corrections have recently been studied, including electron
self-interaction loops \cite{Pachucki:2004si,Pachucki:2005px}, vacuum
polarization \cite{Persson97,DelbrueckKarshenboim,Jentschura:2009at},
and recoil corrections \cite{RecoilShabaev,RecoilShabYerAllZalpha}.
Some effects have been determined to all orders in $Z\alpha$ 
\cite{YerJen2008,Yerokhin:2002pt,yerokhin:04:pra,UehlingKarshenboim2001}
while for others a few terms of the expansion around small $Z\alpha$
have been calculated. 

Here we take the nucleus to be infinitely heavy and neglect small
hadronic and weak effects.  The theoretical
prediction for $g$ in this limit can be organized as a double series
expansion in $\alpha/\pi$ (describing electron selfinteraction) and
$Z\alpha$, describing interactions between the electron and the
nucleus, i.e. the binding effects.  

If we neglect selfinteractions, $g$ is given by Breit's result
(\ref{eq:gCoulomb}).  On the other hand, if we neglect the binding,
$g$ is the same as for a free electron, $g=2+2a$, with the anomaly $a$
described by eq.~(\ref{eq:g-2}).

The interplay of the two types of effects occurs first at the order
$\order{{\alpha\over \pi}(Z\alpha)^2}$.  Interestingly, at this order
in $Z\alpha$, the correction is universal to all orders in $\alpha$.
It is found by considering the magnetic interaction of a bound
particle, taking into account its anomalous magnetic moment.  The
result is  \cite{grotch70,Czarnecki:2000uu}
$\delta g=2a\left(1 + {(Z\alpha)^2 \over 6} \right)$,
where $a$ is given in (\ref{eq:g-2}).

There are no effects linear or cubic in $Z\alpha$, but beginning with
$(Z\alpha)^4$ all higher powers are present.  In fact, in the
approximation of a single-photon electron selfinteraction, it is
possible to use the exact form of the electron propagator in the
Coulomb field and numerically determine effects of
$\order{{\alpha\over \pi}(Z\alpha)^n}$ to all orders $n$
\cite{YerJen2008,Yerokhin:2002pt,yerokhin:04:pra}. 

It is still very
interesting to compute analytically the coefficients of the $Z\alpha$
expansion to check the numerical evaluation.  Such a calculation was
performed for the terms ${\alpha\over \pi}(Z\alpha)^4$ and
${\alpha\over \pi}(Z\alpha)^4\ln Z\alpha$ only in 2004
\cite{Pachucki:2004si}, which illustrates the difficulty of performing
even one-loop calculations for bound particles.  

The situation becomes even harder when two selfinteraction loops are
included.  In this case no numerical all-order study has been
performed.  The terms $\left({\alpha\over \pi}\right)^2(Z\alpha)^4$ and
$\left({\alpha\over \pi}\right)^2(Z\alpha)^4\ln Z\alpha$ are already
known \cite{Pachucki:2005px}.  Together with measurements with
carbon ions \cite{Haffner00} the resulting theoretical prediction
provides the best value of the electron-to-proton mass ratio,
\begin{equation}
{m_p\over m_e}=1836.152\,672\,9(10)
\label{eq:ptoe}
\end{equation}
where we have used the CODATA recommended value of the proton mass in
atomic units,
$m_p =1.007 276 466 77(10) u$, which contributes an order of magnitude
less to the uncertainty in (\ref{eq:ptoe}) than the electron.  We see
that the error in (\ref{eq:ptoe}) is four times smaller than in the
determination using a free electron, eq.~(\ref{eq:vanDyck}).
 
The uncertainty in (\ref{eq:ptoe}) is dominated by the experiment; the
theory is estimated to contribute only about three per cent of the
total error.  The reliability of the theoretical prediction is based
on the numerical knowledge of higher-order binding effects in the
one-loop selfinteraction.  

It is thus very important to independently check the numerical
studies by computing further terms in the $Z\alpha$ expansion.  In
particular, the coefficient of the contribution
${\alpha\over \pi}(Z\alpha)^5$ to $g$ (for the 
principal quantum number $n=1$, that is the ground state) is estimated
to be $H_1 (Z=0)=23.15(10)$ \cite{YerokhinPrivate}, in the notation of
\cite{YerJen2008}. 

The two-loop effect $\left({\alpha\over \pi}\right)^2(Z\alpha)^5$ is
estimated to have the coefficient between $-118$ and $-75$
\cite{Jentschura:2009at}.  

In order to verify these predictions, we have undertaken an
explicit determination of the coefficients of $(Z\alpha)^5$ terms at
one and two-loop selfinteraction level.  In the next section we
present some details of our method with examples of simpler
observables, not involving an external magnetic field: the Lamb shift
in hydrogen-like ions, and the hyperfine splitting (HFS) in muonium.

\section{Lamb shift and hyperfine splitting\label{sec::method}}

\begin{figure}[ht]
  \includegraphics[width=\columnwidth]{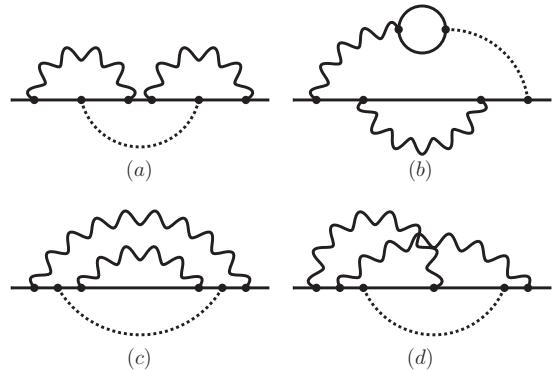}
  \caption{\label{fig::dias} Sample diagrams for the Lamb shift and
    hyperfine splitting at order $\alpha^2 (Z\alpha)^5$. Solid and wavy
    lines denote electrons and photons. The dotted lines denote the
    interaction with the nucleus (cf. Fig~\ref{fig::delta}).}
\end{figure}

In this section we describe our calculations of the radiative-nonrecoil
corrections to the Lamb shift~\cite{Dowling:2009md} and the hyperfine
splitting~\cite{Mondejar:2010se} of hydrogen-like atoms. In both cases
the shift of the energy levels is given by a delta function potential,
which affects only $S$ states,
\begin{equation}
  \delta E = - \mathcal{M}\, |\psi(0)|^2 = - \mathcal{M}
  \frac{(Z\alpha\mu)^3}{\pi n^3} \,.
\end{equation}
$\psi(0)$ is the wave function of a bound $S$ state with principal
quantum number $n$ and reduced mass $\mu$ at the origin. The amplitude
$\mathcal{M}$ is determined from the corresponding Feynman
diagrams. Sample diagrams are shown in Fig.~\ref{fig::dias}.

\begin{figure}[ht]
  \includegraphics[width=\columnwidth]{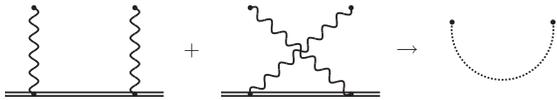}
  \caption{\label{fig::delta} The interaction with the nucleus (double
    line) consists of the sum of direct and crossed photon exchange.}
\end{figure}
The interaction of the electron with the nucleus is depicted in
Fig.~\ref{fig::delta}. In our calculation we construct an expansion in
the ratio of electron and nucleus mass, using the method of expansion by
regions~\cite{Smirnov:2002pj}. We are only interested in the leading
term, which is given by the region where all loop momenta are of the
order of the electron mass. Expanding the nucleus propagators we find
that to leading order the only difference between the diagrams with
direct and crossed photon exchange is an overall sign and the sign of
the $i\varepsilon$ prescription of the latter. We have
\begin{eqnarray}
  \lefteqn{\frac{1}{(q + k)^2 - M^2 + i\varepsilon} +
    \frac{1}{(q - k)^2 - M^2 + i\varepsilon}} && \nonumber\\
  & \to & \frac{1}{2q\cdot k + i\varepsilon} -
  \frac{1}{2q\cdot k - i\varepsilon} + \dots \nonumber \\
  & = & -2i\pi\, \delta( 2q\cdot k) + \dots \,,
\end{eqnarray}
where the ellipses stand for higher order terms. $q$ and $M$ are the
momentum and mass of the nucleus with $q^2 = M^2$, and $k$ is the loop
momentum. In the rest frame of the nucleus, the factor $\delta(
2q\cdot k)=\delta(k_0)/2M$ ensures that the photons exchanged between
the electron and the nucleus carry no energy.  The integration over
the loop denoted with the dotted line in the Feynman diagrams in the
Figures is only over the spatial components $\vec{k}$.

Our calculation is automated to a large extent. We use the program
{\sc qgraf}~\cite{Nogueira:1991ex} to generate all Feynman diagrams.
These are turned into {\sc form}-readable~\cite{Vermaseren:2000nd}
expressions by {\sc q2e} and {\sc
  exp}~\cite{Harlander:1997zb,Seidensticker:1999bb}. The program {\sc
  matad3}~\cite{Steinhauser:2000ry} is used to perform the traces and
express the amplitude $\mathcal{M}$ in terms of scalar integrals,
using custom made routines. Finally, we use integration-by-parts
identities~\cite{Tkachov:1981wb,Chetyrkin:1981qh} to express all
scalar integrals in terms of 32 so-called master integrals. This is
achieved with the help of the program {\sc
  fire}~\cite{Smirnov:2008iw}, which implements the so-called Laporta
algorithm~\cite{Laporta:1996mq,Laporta:2001dd}. Results for all master
integrals and details of their calculation are given in
Ref.~\cite{Dowling:2009md}.

\section{Conclusions}

Our results for the Lamb shift and hyperfine splitting at order
$\alpha^2 (Z\alpha)^5$ are
\begin{eqnarray}
  \delta E_{\mathrm{Lamb}} &=& -6.86100(4)\, \frac{\alpha^2
    (Z\alpha)^5}{\pi n^3} \left( \frac{\mu}{m_e}\right)^3 m_e\,,
  \nonumber \\
  && \\
  \delta E_{\mathrm{HFS}} &=& 0.77099(2)\, \frac{\alpha^2 (Z\alpha)}{\pi
    n^3} E_F \,,
\end{eqnarray}
respectively. $E_F = (8 \mu^3 (Z\alpha)^4 g_N)/(6 m_e M)$ is the Fermi
energy. $g_N$ is the gyromagnetic factor of the nucleus. Our results
improve the precision of the previous results in~\cite{Eides:1995ey} by
more than an order of magnitude. Results for individual diagrams,
including new analytical results, are given in
Refs.~\cite{Dowling:2009md,Mondejar:2010se}.

Our new results agree very well with \cite{Eides:1995ey}, which gives
us confidence in the values of master integrals we have derived.  The
same set of 32 integrals is sufficient to compute the bound electron
$g$ factor to the desired accuracy, $\left({\alpha\over \pi}\right)^2
(Z\alpha)^5$. The only remaining obstacles are the larger number of
diagrams and the necessity of determining corrections  to the wave
function of the ion.  

As the first step, we have found the gauge invariant set of one-loop
(vertex) corrections to the electron interaction with the magnetic
field, an ingredient of the coefficient of ${\alpha\over \pi}
(Z\alpha)^5$, providing approximately a quarter of the estimated value
of 23.15(10) \cite{YerokhinPrivate}.  Work on the remaining one and
two-loop diagrams is in progress.

\subsection*{Acknowledgments }

This research was supported by the Science and Engineering Research
Canada and the Alberta Ingenuity Foundation.

%\bibliographystyle{h-elsevier2}
%\bibliography{/cygdrive/d/pro/Tables/Archive/phd}

\begin{thebibliography}{10}

\bibitem{Brown:1985rh}
L.S. Brown and G. Gabrielse,
\newblock Rev. Mod. Phys. 58 (1986) 233.
%%CITATION = RMPHA,58,233;%%

\bibitem{kinoshitaLeptMomBook}
T. Kinoshita,
\newblock Lepton $g-2$ from 1947 to present,
\newblock Lepton Dipole Moments, edited by B.L. Roberts and W.J. Marciano, 
  Adv. Ser. Dir. HEP Vol.~20, p.~69, World Scientific, Singapore, 2009.
%%CITATION = NONE;%%

\bibitem{laportaLeptMomBook}
S. Laporta and E. Remiddi,
\newblock Analytic {QED} calculations of the anomalous magnetic moment of the
  electron,
\newblock Lepton Dipole Moments, edited by B.L. Roberts and W.J. Marciano,
  Adv. Ser. Dir. HEP Vol.~20, p. 119, World Scientific, Singapore, 2009.
%%CITATION = NONE;%%

\bibitem{Hanneke:2008tm}
D. Hanneke, S. Fogwell and G. Gabrielse,
\newblock Phys. Rev. Lett. 100 (2008) 120801, arXiv:0801.1134.
%%CITATION = PRLTA,100,120801;%%

\bibitem{vanDyck}
D.L. Farnham, R.S. {Van Dyck, Jr.} and P.B. Schwinberg,
\newblock Phys. Rev. Lett. 75 (1995) 3598.
%%CITATION = PRLTA,75,3598;%%

\bibitem{Mohr:2008zz}
P.J. Mohr, B.N. Taylor and D.B. Newell,
\newblock Rev. Mod. Phys. 80 (2008) 633.
%%CITATION = RMPHA,80,633;%%

\bibitem{Cadoret:2008st}
M. Cadoret et~al.,
\newblock Phys. Rev. Lett. 101 (2008) 230801, arXiv:0810.3152.
%%CITATION = PRLTA,101,230801;%%

\bibitem{Wicht2002}
A. Wicht et~al.,
\newblock Phys. Scr. T102 (2002) 82.
%%CITATION = NONE;%%

\bibitem{Guellati2010}
S. Guellati-Khelifa,
%\newblock Determination of the fine structure constant using Bloch oscillations
%  in an accelerated optical lattice,
\newblock talk at the Intl. Conf. on Precision Physics of Simple Atomic
  Systems, Les Houches, June 2010.
%%CITATION = NONE;%%

\bibitem{jeffery97}
A.M. Jeffery et~al.,
\newblock IEEE Inst. Meas. 46 (1997) 264.
%%CITATION = NONE;%%

\bibitem{PhysRevA.79.060503}
J.S. Borbely et~al.,
\newblock Phys. Rev. A 79 (2009) 060503.
%%CITATION = PHRVA,A79,060503;%%

\bibitem{PhysRevLett.104.070403}
K. Pachucki and V.A. Yerokhin,
\newblock Phys. Rev. Lett. 104 (2010) 070403.
%%CITATION = PRLTA,104,070403;%%

\bibitem{PhysRevLett.36.1488}
W.H. Wing et~al.,
\newblock Phys. Rev. Lett. 36 (1976) 1488.
%%CITATION = PRLTA,36,1488;%%

\bibitem{PhysRevA.79.012501}
V.I. Korobov, L. Hilico and J.P. Karr,
\newblock Phys. Rev. A 79 (2009) 012501.
%%CITATION = PHRVA,A79,012501;%%

\bibitem{schneiderl2010}
T. Schneider et~al.,
\newblock Nature Phys. 6 (2010) 275.
%%CITATION = NPAHA,6,275;%%

\bibitem{Breit28}
G. Breit,
\newblock Nature 122 (1928) 649.
%%CITATION = NATUA,122,649;%%

\bibitem{Dirac:1928hu}
P.A.M. Dirac,
\newblock Proc. Roy. Soc. Lond. A117 (1928) 610.
%%CITATION = PRSLA,A117,610;%%

\bibitem{Schwinger:II}
J. Schwinger,
\newblock Particles, sources and fields, vol. II (Addison-Wesley, Redwood City,
  {CA}, 1973).
%%CITATION = NONE;%%

\bibitem{Pachucki:2004si}
K. Pachucki, U.D. Jentschura and V.A. Yerokhin,
\newblock Phys. Rev. Lett. 93 (2004) 150401, hep-ph/0411084,
\newblock Erratum Phys. Rev. Lett. {\bf 94}, 229902 (2005).
%%CITATION = PRLTA,93,150401;%%

\bibitem{Pachucki:2005px}
K. Pachucki et~al.,
\newblock Phys. Rev. A 72 (2005) 022108, physics/0506227.
%%CITATION = PHRVA,A72,022108;%%

\bibitem{Persson97}
H. Persson et~al.,
\newblock Phys. Rev. A56 (1997) R2499.
%%CITATION = PHRVA,A56,2499;%%

\bibitem{DelbrueckKarshenboim}
S.G. Karshenboim and A.I. Milstein,
\newblock Phys. Lett. B 549 (2002) 321.
%%CITATION = PHLTA,B549,321;%%

\bibitem{Jentschura:2009at}
U.D. Jentschura,
\newblock Phys. Rev. A 79 (2009) 044501.
%%CITATION = PHRVA,A79,044501;%%

\bibitem{RecoilShabaev}
V.M. Shabaev,
\newblock Phys. Rev. A 64 (2001) 052104.
%%CITATION = PHRVA,A64,052104;%%

\bibitem{RecoilShabYerAllZalpha}
V.M. Shabaev and V.A. Yerokhin,
\newblock Phys. Rev. Lett. 88 (2002) 091801.
%%CITATION = PRLTA,88,091801;%%

\bibitem{YerJen2008}
V.A. Yerokhin and U.D. Jentschura,
\newblock Phys. Rev. Lett. 100 (2008) 163001.
%%CITATION = PRLTA,100,163001;%%

\bibitem{Yerokhin:2002pt}
V.A. Yerokhin, P. Indelicato and V.M. Shabaev,
\newblock Phys. Rev. Lett. 89 (2002) 143001, hep-ph/0205245.
%%CITATION = PRLTA,89,143001;%%

\bibitem{yerokhin:04:pra}
V.A. Yerokhin, P. Indelicato and V.M. Shabaev,
\newblock Phys. Rev. A 69 (2004) 052503.
%%CITATION = PHRVA,A69,052503;%%

\bibitem{UehlingKarshenboim2001}
S. Karshenboim, V. Ivanov and V. Shabaev,
\newblock Can. J. Phys. 79 (2001) 81.
%%CITATION = CJPHA,79,81;%%

\bibitem{grotch70}
H. Grotch,
\newblock Phys. Rev. Lett. 24 (1970) 39.
%%CITATION = PRLTA,24,39;%%

\bibitem{Czarnecki:2000uu}
A. Czarnecki, K. Melnikov and A. Yelkhovsky,
\newblock Phys. Rev. A63 (2001) 012509, hep-ph/0007217.
%%CITATION = PHRVA,A63,012509;%%

\bibitem{Haffner00}
H. {H\"affner} et~al.,
\newblock Phys. Rev. Lett. 85 (2000) 5308.
%%CITATION = PRLTA,85,5308;%%

\bibitem{YerokhinPrivate}
V.A. Yerokhin, 2010,
\newblock private communication.
%%CITATION = NONE;%%

\bibitem{Dowling:2009md}
M. Dowling et~al.,
\newblock Phys. Rev. A81 (2010) 022509, arXiv:0911.4078.
%%CITATION = PHRVA,A81,022509;%%

\bibitem{Mondejar:2010se}
J. Mond\'ejar, J.H. Piclum and A. Czarnecki,
\newblock Phys. Rev. A81 (2010) 062511, arXiv:1005.1944.
%%CITATION = 1005.1944;%%

\bibitem{Smirnov:2002pj}
V.A. Smirnov,
\newblock Springer Tracts Mod. Phys. 177 (2002) 1.
%%CITATION = STPHB,177,1;%%

\bibitem{Nogueira:1991ex}
P. Nogueira,
\newblock J. Comput. Phys. 105 (1993) 279.
%%CITATION = JCTPA,105,279;%%

\bibitem{Vermaseren:2000nd}
J.A.M. Vermaseren,
\newblock New features of FORM,
\newblock math-ph/0010025, 2000.
%%CITATION = MATH-PH/0010025;%%

\bibitem{Harlander:1997zb}
R. Harlander, T. Seidensticker and M. Steinhauser,
\newblock Phys. Lett. B426 (1998) 125, hep-ph/9712228.
%%CITATION = PHLTA,B426,125;%%

\bibitem{Seidensticker:1999bb}
T. Seidensticker,
%\newblock Automatic application of successive asymptotic expansions of Feynman
%  diagrams,
\newblock hep-ph/9905298, 1999.
%%CITATION = HEP-PH/9905298;%%

\bibitem{Steinhauser:2000ry}
M. Steinhauser,
\newblock Comput. Phys. Commun. 134 (2001) 335, hep-ph/0009029.
%%CITATION = CPHCB,134,335;%%

\bibitem{Tkachov:1981wb}
F.V. Tkachov,
\newblock Phys. Lett. B100 (1981) 65.
%%CITATION = PHLTA,B100,65;%%

\bibitem{Chetyrkin:1981qh}
K.G. Chetyrkin and F.V. Tkachev,
\newblock Nucl. Phys. B192 (1981) 159.
%%CITATION = NUPHA,B192,159;%%

\bibitem{Smirnov:2008iw}
A.V. Smirnov,
%% \newblock {Algorithm FIRE -- Feynman Integral REduction},
\newblock JHEP 0810 (2008) 107, arXiv:0807.3243.
%%CITATION = JHEPA,0810,107;%%

\bibitem{Laporta:1996mq}
S. Laporta and E. Remiddi,
\newblock Phys. Lett. B379 (1996) 283, hep-ph/9602417.
%%CITATION = PHLTA,B379,283;%%

\bibitem{Laporta:2001dd}
S. Laporta,
\newblock Int. J. Mod. Phys. A15 (2000) 5087, hep-ph/0102033.
%%CITATION = IMPAE,A15,5087;%%

\bibitem{Eides:1995ey}
M.I. Eides and V.A. Shelyuto,
\newblock Phys. Rev. A52 (1995) 954, hep-ph/9501303.
%%CITATION = PHRVA,A52,954;%%

\end{thebibliography}
\end{document}